\documentclass[letterpaper, 10 pt, conference]{ieeeconf}

\usepackage{graphicx}
\usepackage{mathtools}
\usepackage{url}
\usepackage{xcolor}
\usepackage[ruled,vlined]{algorithm2e}

\usepackage[utf8]{inputenc}

\IEEEoverridecommandlockouts                              

\title{\LARGE \bf
The Impact of Ridesharing in Mobility-on-Demand Systems: Simulation Case Study in Prague}

\author{Davide Fiedler$^{1}$, Michal \v{C}ertick\'{y}$^{1}$, Javier Alonso-Mora$^{2}$, and Michal \v{C}\'{a}p$^{1,2}$
\thanks{$^{1}$Department of Computer Science, Faculty of Electrical Engineering, CTU in Prague, Czech Republic
        {\tt\scriptsize david.fiedler@agents.fel.cvut.cz}}%
\thanks{$^{2}$Department of Cognitive Robotics, 3ME, TU Delft, the Netherlands}%
}

\begin{document}

\maketitle
\thispagestyle{empty}
\pagestyle{empty}

\begin{abstract}

In densely  populated-cities, the use of private cars for personal transportation is unsustainable, due to high parking and road capacity requirements. 
The mobility-on-demand systems have been proposed as an alternative to a private car.
Such systems consist of a fleet of vehicles that the user of the system can hail for one-way point-to-point trips. 
These systems employ large-scale vehicle sharing, i.e., one vehicle can be used by several people during one day and consequently the fleet size and the parking space requirements can be reduced, but, at the cost of a non-negligible increase in vehicles miles driven in the system. 
The miles driven in the system can be reduced by ridesharing, where several people traveling in a similar direction are matched and travel in one vehicle. 
We quantify the potential of ridesharing in a hypothetical mobility-on-demand system designed to serve all trips that are currently realized by private car in the city of Prague.
Our results show that by employing a ridesharing strategy that guarantees travel time prolongation of no more than 10 minutes, the average occupancy of a vehicle will increase to 2.7 passengers. 
Consequently, the number of vehicle miles traveled will decrease to 35 \% of the amount in the MoD system without ridesharing and to 60\% of the amount in the present state.

\end{abstract}

\section{Introduction}

In densely-populated cities, private cars are an unsustainable mode of personal transportation. Parking capacity and road capacity in such environments are typically insufficient to accommodate the private car traffic. At the same time, roads and parking space are difficult to expand due to scarcity of free land. In result, many modern cities are plagued by congested roads, unavailability of parking, and high levels of air pollution.

Mobility-on-demand (MoD) systems represent an alternative to transportation by private vehicles. 
In an MoD system, when a passenger requests a transportation service, a particular vehicle is assigned to pick up the passenger and carry him or her to the desired destination, where
the passenger is dropped off. Then, the vehicle is available again to serve other passengers. When an on-demand vehicle travels without a passenger, e.g., when it is moving from one drop-off position to
the next pick up position, we say that it drives unallocated. 
Examples of on-demand systems include taxi service, transportation network companies such as Uber, or future systems of self-driving taxis being developed by companies such as Waymo, Uber or nuTonomy. The MoD systems employ massive {\em vehicle sharing}, and thus they can serve the existing transportation demand with smaller highly-utilized vehicle fleet and in turn, they promise to dramatically reduce the need for urban parking capacity. 
The performance of large-scale mobility-on-demand systems has been recently studied within newly developed mathematical models that indeed confirm the ability to reduce the number of vehicles in the system~\cite{fagnant_travel_2014, burns_transforming_2013}.
However, they also show that the total amount of vehicle miles driven in the system is bound to increase due to unallocated vehicle trips~\cite{spieser_toward_2014}. 
For example, the case studies in Singapore~\cite{spieser_toward_2014} and Prague~\cite{fiedler_impact_2017} indicate that unallocated traffic in the MoD system would add more than 30\,\% to total traffic in the system and subsequently would contribute to the formation of congestion~\cite{fiedler_impact_2017}. 

A possible solution to the above problem is to implement large-scale {\em ridesharing}, where multiple passengers that travel in the same direction are matched and transported in one vehicle. Efficient ridesharing can increase vehicle occupancy and consequently reduce the required fleet size and the total amount of vehicle miles driven in the system. 

The objective of this paper is to quantify the potential of ridesharing to reduce vehicular traffic in a large-scale MoD system. We use agent-based simulation to analyze hypothetical scenario of replacing all privately owned vehicles in the city of Prague by a) a single-occupancy MoD system and by b) an MoD system with ridesharing.

\subsection*{Related work}
The impact of potential large-scale MoD system deployments have been studied extensively in recent years, both using simulation~\cite{fagnant_travel_2014} and by means of formal mathematical modeling of such systems~\cite{spieser_toward_2014}. The impact on congestion arising from unallocated vehicle trips in MoD system has been numerically explored in~\cite{zhang_model_2016} and \cite{fiedler_impact_2017}. One of the proposed mitigation is strategies is congestion-aware fleet routing~\cite{rossi_routing_2018}. 

Ridesharing was traditionally formalized in the framework of Vehicle Routing Problems~(VRP)~\cite{toth_vehicle_2014}, typically as a specific variant of VRP with Pickup and Delivery~\cite{berbeglia_static_2007,berbeglia_dynamic_2010,baldacci_exact_2011,grandinetti_multi-objective_2014}
or a Dial a Ride Problem (DARP)~\cite{cordeau_dial--ride_2007,berbeglia_dynamic_2010}.
Yet, the existing VRP methods focus on instances with tens of vehicles and requests~\cite{parragh_hybrid_2013,mahmoudi_finding_2016} and as such, they are not applicable to management or analysis of large-scale fleets that often consist of thousands of vehicles and requests.

The potential for large-scale ridesharing within MoD systems was studied using the shareability network
model in~\cite{santi_quantifying_2014}. This analysis of taxi trips in Manhattan revealed that
up to 80$\%$ of the trips could be pairwise shared such that the travel time is increased be no more than a couple of minutes. The analysis was later extended to other cities~\cite{tachet_scaling_2017}.
The assumption of maximum two passengers in a vehicle was later lifted in \cite{alonso-mora_-demand_2017}, where the
authors proposed a technique for dynamic assignment of requests to vehicles in the fleet and apply the technique to analyze the potential of ridesharing within NYC taxi dataset. In contrast to the above work, that analyzed the scenario of replacing all taxis in Manhattan by a fleet of 3\,000 shared taxis, we analyze the scenario of replacing all private vehicles with a fleet of 50\,000 shared MoD vehicles and specifically study how would large-scale ridesharing affect the utilization of road infrastructure.

The city of Prague was chosen for the case study partly because we have access to the demand model for the area and partly because it is a "regular" metropolis that shares many characteristics with other world cities and thus the results are more likely to generalize. This is in contrast with the previously considered urban areas such as Manhattan or Singapore that have an extremely high density of travel demand which may lead to overly-optimistic estimates of system performance. 

\subsection*{Contribution}
Our work extends the existing models of large-scale MoD systems with consideration for the potential of ridesharing. More specifically, using agent-based simulation, we quantify the potential of ridesharing to reduce traffic load and total vehicle miles traveled. We compare how would be the existing travel demand served a) in the current system with private vehicles, b) in the MoD system without ridesharing and c) in the MoD system with ridesharing.  
We run high-fidelity realistic-scale simulation of the three scenarios for the city of Prague. In particular, we use a) real road network topology, b) demand with realistic structure and realistic scale, and c) we simulate the whole lifetime of the vehicle, including empty trips.

Analyzing of the ridesharing in context of on-demand systems is essential because it can answer the question of whether the MoD system can be, in fact, deployed without overloading the capacity of existing road infrastructure.

\section{Background Material}
When analyzing the traffic congestion on a road segment, the  \emph{fundamental diagram of road traffic}~\cite{tanner_mathematical_1964} is often used. 
This model relates traffic density to traffic flow.
\emph{Traffic density} is the number of vehicles per unit of distance on the road segment, while \emph{traffic flow} is the number of vehicles passing a reference point per unit of time.
When traffic density grows, traffic flow also increases until it reaches a tipping point after which the flow starts dropping due to congestion.
The tipping point is known as the critical density~\cite{kerner_introduction_2009}.
The exact shape of the diagram is typically determined by fitting empirical data from real-world observations of vehicular traffic.
Subsequently, we will use the traffic density as a measure of utilization of a particular road segment and the critical density value $ y_c = 0.08 \,\mathrm{vehicle}\, \mathrm{m}^{-1} $  from~\cite{tadaki_critical_2015}. 
The road segments with traffic density above critical density \( y_c \) will be referred to as congested road segments.

\section{Methodology}
In this section, we describe the methodology that allows us to quantify the potential of large-scale ridesharing in a hypothetical scenario of replacing all private cars in the city of Prague by an MoD system. 

First, we use a travel demand model to generate a representative collection of trips that are currently realised by private cars in Prague during the morning peak. Then, we design an MoD system that is capable of serving the existing trips with required service quality. Further, we design a method that finds request-vehicle matching for ridesharing. Finally, we simulate the entire system in micro-simulation and gather data for subsequent analysis. We will now discuss each step in detail. 

\subsection{Input data}

The set of trips that represent the transportation demand is generated by the multi-agent activity-based model of Prague and Central Bohemian Region~\cite{vcerticky2015fully}. In contrast to traditional four-step demand models~\cite{hensher2007handbook}, which use trips as the fundamental modeling unit, activity-based models employ so-called activities (e.g. work, shop, sleep) and their sequences to represent the transport-related behavior of the population. Travel demand then occurs due to the necessity of the agents to satisfy their needs through activities performed at different places at different times. These activities are arranged in time and space into sequential (daily) schedules. Trip origins, destinations and times are endogenous outcomes of activity scheduling. The activity-based approach considers individual trips in context and therefore allows representing realistic trip chains.

The model used in this work covers a typical work day in Prague and the surrounding Central Bohemian Region. The population of over 1.3 million is modeled by the same number of autonomous, self-interested agents, whose behavior is influenced by their sociodemographic attributes, current needs, and situational context. Individual decisions of the agents are implemented using machine learning methods (neural networks, decision trees, random forests, etc.) and trained using various real-world data sets, including census data, travel diaries, and other transportation-related surveys. Planned activity schedules are simulated and tuned, and, finally, their temporal, spatial, and structural properties are validated against additional historical real-world data (origin-destination matrices and surveys) using six-step validation framework VALFRAM~\cite{jass2016, drchal2015data}. 
The model generates over three million trips by all modes of transport in one 24-hour scenario, out of which there are roughly one million trips by private vehicles. 
For the analysis, we select only the personal car trips during the morning peak, i.e, such that start between 06:30 and 08:00, which yields a collection of about 130 000 trips.

\subsection{System model}
\label{sec:system_model}

\begin{figure}
\centering{}\includegraphics[width=1\columnwidth]{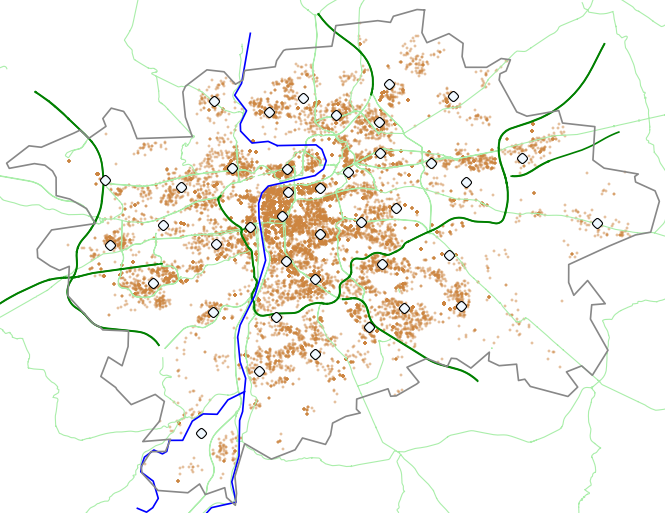}\caption{\label{fig:stations}MoD system stations in the city of Prague. Stations are shown as circles. Spatial distribution of origins and destinations of travel demands is depicted using brown dots.}
\end{figure}

We adopt a station-based design of the MoD system~\cite{spieser_toward_2014}. That is, we partition the city into $n=40$ regions using k-means clustering over the demand data, and we assume that there is a station at the center of each such region. The result of this process is shown in Figure~\ref{fig:stations}. The number of regions was chosen such that the average travel time from station to a passenger is below 3 minutes. Stations serve as temporary parking lots for idle vehicles, and they also contain facilities such as refueling/charging and cleaning. 

For simplicity, we assume that each station has sufficient stock of vehicles to cover the transportation demand originating in its region. Further, we assume that the stock of vehicles at each station is stabilized through a vehicle rebalancing process, that is, the stations with a surplus of idle vehicles continuously send empty vehicles to stations that have a shortage of idle vehicles. In particular, we use the rebalancing policy based on the solution of the optimal transport problem~\cite{pavone_robotic_2012} used, e.g., in the Singapore MoD case study~\cite{spieser_toward_2014}. 

We experimentally determined that in order to be able to serve every request from the nearest station (without ridesharing), the MoD system requires the total of 51 951 vehicles.

The MoD system is modeled as follows. Let $V=1,\ldots,m$ be the set of all vehicles in the system. The transportation requests arriving to the system are represented by a sequence $(t_1, o_1, d_1), (t_2, o_2, d_2), \ldots \ $, where $t_i, o_i,$ and $d_i$ are the announcement time, origin point, and destination point of request $i$ respectively. The $i$-th transportation request is revealed only at time $t_i$.

The state of a vehicle $v$ at a particular time point encodes its current position, the set of requests that are currently on-board of the vehicle and the current plan of the vehicle. The plan of a vehicle is represented as a sequence of plan orders $\pi = z_1,z_2,\ldots \ $, where each order $z_{i}$ is either to pick up a request~$r$, or to drop-off a request~$r$. For a vehicle plan to be valid, for each on-board passenger, the plan must contain an order to drop-off this passenger and for every order to pickup request $r$, the plan must contain an order to drop-off the request $r$ later in the sequence. 

The operational cost of vehicle $v$ when following plan $\pi$ is denoted $s_v(\pi)$. For simplicity, we define $s_v(\pi)$ to be equal to the distance driven by the vehicle when it follows plan~$\pi$. The travel delay (or discomfort) of request $r$ when the request is served by vehicle $v$ following plan $\pi_v$ is denoted $q_r(\pi_v)$. We define the travel delay as the difference between the travel time in a shared vehicle and a travel time along the fastest route: $$q_r(\pi_v) := (t_r^\mathrm{dropoff} - t_r) - t^\mathrm{baseline}_r,$$ where $t_r^\mathrm{dropoff}$ is the time when the request was dropped off under plan $\pi_v$ and $t^\mathrm{baseline}_r$ is the duration of the fastest route from pickup point of request $r$ to the drop-off point of request $r$.  

We desire to minimize the total operational cost of the system, such that the discomfort of every passenger is bounded by $q_\mathrm{max}$. That is, we will attempt to minimize 
$$ \sum_v s_v(\pi_v) $$ 
subject to 
$$ q_r(\pi_{v(r)}) \leq q_\mathrm{max} \quad \forall r ,$$
where $\pi_{v(r)}$ is the plan of the vehicle that serves the request~$r$.

\subsection{Request-vehicle matching}
\label{sec:matching}
To implement ridesharing in an MoD system, one has to determine how should be the arriving travel requests assigned to individual vehicles in the system such that the overall distance driven in the system is minimized. 
This problem can be mathematically modeled as a dynamic dial-a-ride problem (D-DARP), which is known to be NP-hard~\cite{parragh_survey_2008}. Despite recent algorithmic advancements~\cite{ropke_branch_2009,alonso-mora_-demand_2017,cap_multi-objective_2018}, the scalability of exact methods remains limited to problem instances of moderate size. 
Therefore, in this work, we compute distance-minimizing request-vehicle ridesharing assignments using an \emph{Insertion Heuristic}~\cite{jung_dynamic_2015}. The request-vehicle matching is implemented as follows:

When a new request $r=(t,o,d)$ arrives, the request-vehicle matching algorithm attempts to incorporate the request into the current plan of every vehicle. For a particular vehicle $v$, we try all possible points to insert pickup and drop-off order into existing plan, such that the relative ordering of the previous orders in the plan remains unchanged. 
Further, we measure the increase of operational cost for each plan generated by this process and select the plan (and the corresponding vehicle) that minimizes the increase in operational cost and at the same time satisfies the service discomfort constraints. 
The maximum discomfort threshold $q_\mathrm{max}$ is chosen such that, if no other suitable ridesharing match is found, the request can always be served with delay smaller than $q_\mathrm{max}$ using a vehicle dispatched from the nearest system station.
For the pseudo-code of the request-vehicle matching method, see Algorithm~\ref{alg:inseriton_heuristic}. 

In order to evaluate the operational cost and service discomfort associated with a particular plan, one needs to estimate the travel time of the vehicle from one order in the plan (e.g., from pickup position of one request) to the subsequent order in the plan (e.g., the drop-off position of some other request). Notice that the number of evaluation of travel time estimate can be up to the order of $O(m \cdot l^3)$. The fleet size in our case study will be $m \approx 50\,000$ and the average length of the plan will be $l \approx 4$ resulting hundreds of thousands of evaluations of the travel time estimate within single request assignment procedure. In order to maintain computational tractability, instead of planing the fastest route on the road network, we approximate the travel time between two locations on the map with a linear model based on the Euclidean distance between the two points. The model is calibrated to minimize the difference between the estimate and the duration of the fastest route in the road network. Since the road network travel time may be longer than the estimate, a plan that satisfies discomfort constraints computed using estimate may violate the constraint when evaluated using road network travel time. Nevertheless, in our experiments, we observed that the average road network travel delay is significantly lower than the maximum delay constraint.

\SetKwProg{alg}{Algorithm}{}{}
\SetKwProg{on}{On}{}{}
\SetKwProg{function}{Function}{}{}

\SetKwFunction{pp}{PP}
\LinesNumbered

\begin{algorithm}[t]

\on{arrival of new request $r$} {
	\For{$v\in V$}{		
			    let $\pi_v$ be the current plan of vehicle $v$\;
    			let $l$ be the length of plan $p_v$\;
				\For{$i \in {1, \ldots, l + 1}$}{
					\For{$j \in {i + 1, \ldots, l + 2}$}{
		   				$\pi' \leftarrow $ insert \texttt{pick-up($r$)} order before position $i$ in plan $\pi_v$\;
		   				$\pi^{ij}_v \leftarrow $ insert \texttt{drop-off($r$)} order before position $j$ in plan $\pi'$\;
					}	
				} 
	
			$v^*, i^*, j^* \leftarrow \underset{v,i,j}{\mathrm{argmin}} \ s_v(\pi^{i,j}_v) - s_v(\pi_v)$ subj. to $\pi_v^{i,j}$ is valid and $q_r(\pi_v^{ij}) \leq q_\mathrm{max} \ \ \forall \text{ req. }r \text{ served by plan } \pi_v^{ij}$\;
	}
	
	request $r$ is assigned to vehicle $v^*$\;
	vehicle $v^*$ follows plan $p^{i^*j^*}_{v^*}$\;
	
}

\caption{\label{alg:inseriton_heuristic} Finding a vehicle and a plan to serve request~$r$ using insertion heuristic. }
\end{algorithm}

\subsection{Simulation}
\begin{figure}
\centering{}\includegraphics[width=1\columnwidth]{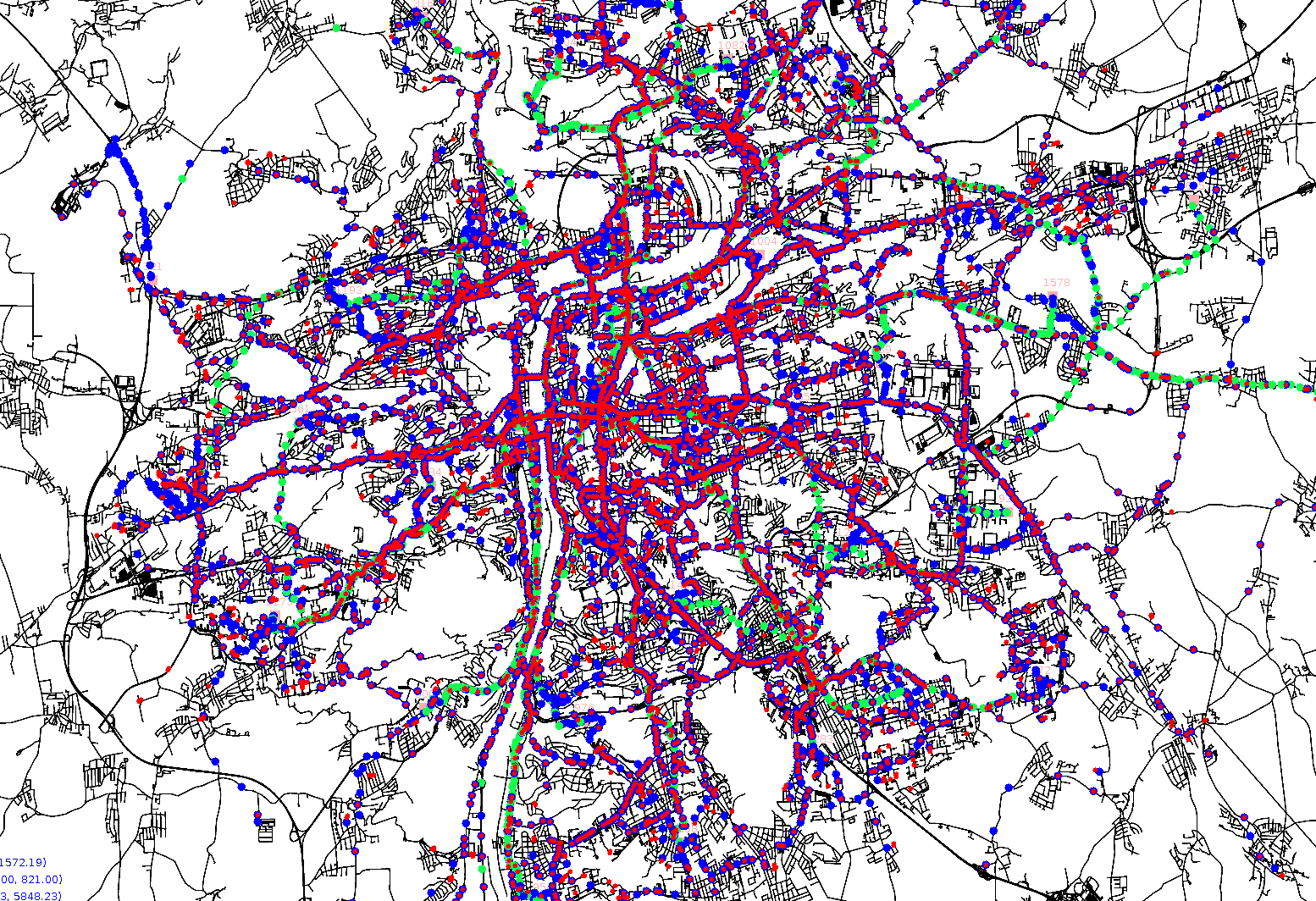}
\begin{tabular}{|c|}
\hline 
\textbf{Video:} \url{https://sum.fel.cvut.cz/itsc2018/}\tabularnewline
\hline 
\end{tabular}
\caption{\label{fig:ap}AgentPolis visualization of the traffic in Prague at 07:00.}
\end{figure}

The scenarios were simulated in multi-agent transportation simulation framework AgentPolis\footnote{\url{https://github.com/aicenter/agentpolis}}. 
AgentPolis is a large-scale multi-agent discrete-event simulation written in Java. 
The simulation environment consists of a) road network that is in turn composed of \emph{nodes} (crossroads) connected by \emph{edges} (road segments), b) stations in which on-demand vehicles park, c) on-demand vehicle agents, and d) passenger agents. In Figure~\ref{fig:ap}, we show the city of Prague simulated in AgentPolis.

The topology of the road network is obtained from OpenStreetMap (OSM), resulting in a road network consisting of 158\,674 edges and 63\,995 nodes.
The speed limit for each road segment was also taken from OpenStreetMap data and missing entries were generated according to following rules based on the local legislation: highway: 130 km/h, living street: 20 km/h, otherwise: 50 km/h. 

During simulation initialization, we create the vehicles representing the on-demand fleet and assign them to stations. 
For each travel request, at the request announcement time, we create a passenger agent. 
The passenger is then picked up by the assigned on-demand vehicle, driven to the desired location, dropped off and finally released from the simulation. 
The vehicle to serve the passenger is selected using the request-vehicle matching procedure described in Section~\ref{sec:matching}. 
Note that each passenger can be either matched to one of the empty vehicles parked in a station or to a non-idle vehicle that is already on its way to serve previously assigned requests. 
After the passenger has been dropped off, the vehicle continues executing its plan. In the case the plan of the vehicle becomes empty, the vehicle returns to the nearest system station. 

In this paper, we analyze the system during the morning peak, i.e., during the period between 7:00 and 8:00. To avoid the ``cold start" artifacts, the simulation begins at 6:30, but for subsequent analysis, we only use the data generated between 7:00 and 8:00.

\section{Results}

In this section, we report on the results of our simulation analysis.
We run the following three scenarios: 
\begin{enumerate}
\item {\em Present situation scenario:} Each travel request is realized by a private vehicle.
\item {\em MoD scenario:} Travel demands is served by the MoD system.
\item {\em MoD with ridesharing scenario:} Travel demand is served by the MoD system with ridesharing. Request-vehicle matching uses maximum travel delay $q_\mathrm{max}=10\,min$.
\end{enumerate}

\begin{figure*}
\centering{}\includegraphics[width=1\linewidth]{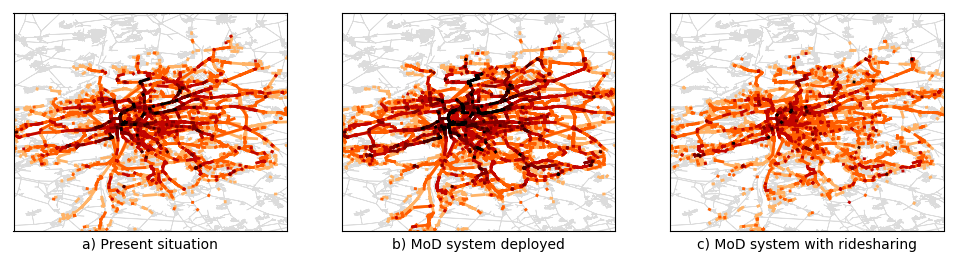}
\caption{\label{fig:traffic_density_map_comparison} Comparison of traffic densities over the road network in the three compared scenarios. The darker colors indicates higher traffic density, black color is reserved for the roads that are congested, i.e., the traffic density exceeds the critical density.}
\end{figure*}

Figure~\ref{fig:traffic_density_map_comparison} shows the distribution of the traffic density over the road network for the three compared scenarios. 
We can see that when the transportation demand is served by an MoD system without ridesharing, the traffic density at most parts of the road network increases compared to the present situation. When the MoD system employs ridesharing, however, the traffic density at most road segments decreases relative to both the present situation and to the MoD system without ridesharing.   

\begin{figure*}
\centering{}\includegraphics[width=1\linewidth]{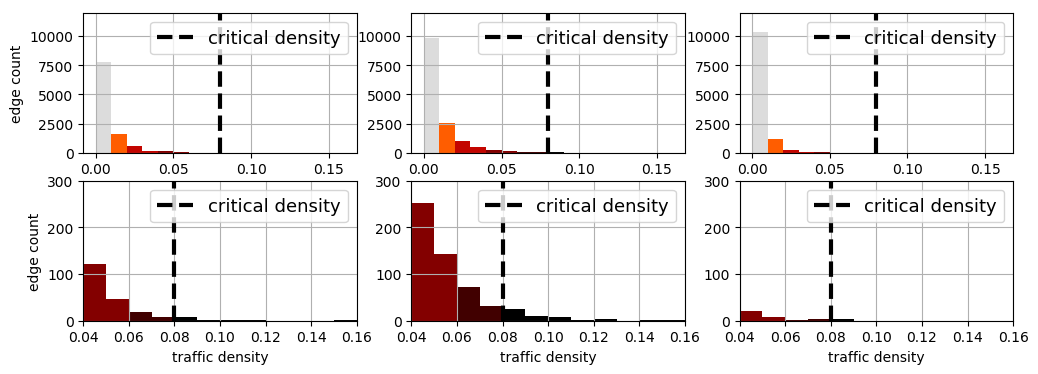}
\caption{\label{fig:traffic_density_histogram_comparison} Histogram of traffic densities in the three different scenarios. In the second row, only the bins with traffic density above 50\% of critical density ($0.04\,\mathrm{vehicle}\,\mathrm{m^{-1}}$) are shown to improve readability. Edges with zero traffic density were discarded from all histograms.} 
\end{figure*}

Figure~\ref{fig:traffic_density_histogram_comparison} shows the histogram of travel densities in the three scenarios. 
The first row shows the traffic density histograms for all edges with non-zero density (used at least once in the analyzed time window) while the second row shows histograms that are ``zoomed-in" to the values around the critical density. 
Again, we can see that MoD system with ridesharing is able to serve the current transportation demand using less road capacity than the private vehicles use today and than the on-demand vehicles would use if they do not employ ridesharing.

Table~\ref{fig:comparison_table} summarizes the performance of the three evaluated systems. It is remarkable that ridesharing can reduce the total amount of vehicle miles traveled in an MoD system (and thus also its energy consumption) almost three-fold.

\begin{table}
\scriptsize
\centering{}%
\begin{tabular}{|c|c|c|c|}
\hline
 & Present & MoD & MoD with ridesharing
\tabularnewline
\hline
\hline
Total veh. dist. traveled (km) & 940 645 & 1 586 495 & 560 875
\tabularnewline
\hline
Avg. dist. traveled/vehicle (km) & 18.1 & 30.5 & 10.8
\tabularnewline
\hline
Avg. density (veh/km) & 0.0080 & 0.0101 & 0.0052
\tabularnewline
\hline
Congested segments & 14 & 55 & 4
\tabularnewline
\hline
Heavily loaded segments & 208 & 551 & 35
\tabularnewline
\hline
\end{tabular}
\caption{\label{fig:comparison_table}Comparison between the three considered scenarios. Congested segments are segments on which traffic density is above critical density, heavily loaded segments are segments with density above 50\% of the critical density.}
\end{table}

\begin{figure}
\centering{}\includegraphics[width=0.75\linewidth]{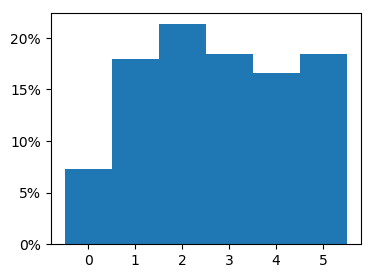}
\caption{\label{fig:occupancy}Vehicle occupancy histogram}
\end{figure}

To understand the effectiveness of ridesharing, the vehicle occupancy is one of the most important indicators.  
The occupancy of each non-idle vehicle (i.e., a vehicle that is not in a station) was measured during the simulation in one-minute intervals.  
Figure~\ref{fig:occupancy} shows a histogram of vehicle occupancies measured in time period 7:00 - 8:00. The MoD system with ridesharing has average occupancy of 2.74 passengers per vehicle. This represents a significant improvement over the MoD system without ridesharing that has the average occupancy of 0.7 passengers per vehicle.

\begin{figure}
\centering{}\includegraphics[width=0.75\linewidth]{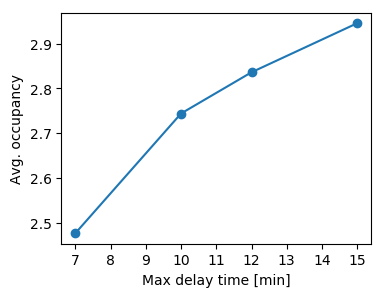}
\caption{\label{fig:occupancy_comparison} Relation between maximum travel delay and the average vehicle occupancy.}
\end{figure}

In order to get better insights into the relationship between the travel discomfort and vehicle occupancy, we simulated MoD system with ridesharing and travel delay constraints $q_\mathrm{max}$ =  7, 10, 12 and 15 minutes.
Figure~\ref{fig:occupancy_comparison} shows the relation between the maximum delay and the average vehicle occupation. As expected, the vehicle occupancy can be increased, and in turn operation cost reduced, if we sacrifice the comfort of passengers.

\section{Conclusion}
In cities, the private vehicle is becoming an unsustainable mode of personal transportation due to its significant parking and road capacity requirements. Mobility-on-Demand (MoD) systems represent an alternative to transportation by private vehicles. These systems employ massive vehicle sharing, and thus they can serve existing transportation demand with a smaller highly-utilized fleet and in turn drastically reduce the need for parking capacity. However, studies show that unallocated trips in such systems would significantly increase vehicular traffic compared to the present state.

A proposed solution to this problem is to employ large-scale ridesharing. The potential of this solution strategy however remained unclear. 

In this paper, we quantified the potential of ridesharing to reduce vehicular traffic in an MoD system. We analyzed hypothetical scenario of MoD deployment in the city of Prague and compared performance of three scenarios: a) the current state, where the demand is served by private vehicles, b) an MoD system without ridesharing and c) an MoD system with ridesharing. 

The simulation results demonstrate that the MoD systems with ridesharing can not only compensate for the traffic generated by the unallocated trips within on-demand systems, but that ridesharing can significantly improve the road network utilization even with respect to the present state. 
In particular, by employing ridesharing, the average total traveled distance was reduced to 35\% of the distance traveled in MoD system without ridesharing and to 60\% of the distance traveled by private vehicles in the present state.

In future, we plan to study the applicability of more sophisticated request-vehicle matching methods to such large-scale scenarios and conduct test with more complex congestion models.

\bibliographystyle{unsrt}
\bibliography{Zotero,michal-certicky,mcap_zotero}

\begin{thebibliography}{10}

\bibitem{fagnant_travel_2014}
Daniel~J. Fagnant and Kara~M. Kockelman.
\newblock The travel and environmental implications of shared autonomous
  vehicles, using agent-based model scenarios.
\newblock {\em Transportation Research Part C: Emerging Technologies},
  40:1--13, March 2014.

\bibitem{burns_transforming_2013}
L.~D Burns, W.~C. Jordan, and B.~A. Scarborough.
\newblock Transforming {{Personal Mobility}}.
\newblock Technical report, {Earth Institute, Columbia University}, January
  2013.

\bibitem{spieser_toward_2014}
Kevin Spieser, Kyle Treleaven, Rick Zhang, Emilio Frazzoli, Daniel Morton, and
  Marco Pavone.
\newblock Toward a {{Systematic Approach}} to the {{Design}} and {{Evaluation}}
  of {{Automated Mobility}}-on-{{Demand Systems}}: {{A Case Study}} in
  {{Singapore}}.
\newblock {\em Road Vehicle Automation (Lecture Notes in Mobility)}, April
  2014.

\bibitem{fiedler_impact_2017}
D.~Fiedler, M.~Čáp, and M.~Čertický.
\newblock Impact of mobility-on-demand on traffic congestion:
  {Simulation}-based study.
\newblock In {\em 2017 {IEEE} 20th {International} {Conference} on
  {Intelligent} {Transportation} {Systems} ({ITSC})}, pages 1--6, October 2017.

\bibitem{zhang_model_2016}
R.~Zhang, F.~Rossi, and M.~Pavone.
\newblock Model predictive control of autonomous mobility-on-demand systems.
\newblock In {\em 2016 {IEEE} {International} {Conference} on {Robotics} and
  {Automation} ({ICRA})}, pages 1382--1389, May 2016.

\bibitem{rossi_routing_2018}
Federico Rossi, Rick Zhang, Yousef Hindy, and Marco Pavone.
\newblock Routing autonomous vehicles in congested transportation networks:
  Structural properties and coordination algorithms.
\newblock {\em Autonomous Robots}, pages 1--16, May 2018.

\bibitem{toth_vehicle_2014}
Paolo Toth and Daniele Vigo.
\newblock {\em Vehicle {{Routing}}: {{Problems}}, {{Methods}}, and
  {{Applications}}, {{Second Edition}}}.
\newblock {SIAM}, December 2014.

\bibitem{berbeglia_static_2007}
Gerardo Berbeglia, Jean-Fran{\c c}ois Cordeau, Irina Gribkovskaia, and Gilbert
  Laporte.
\newblock Static pickup and delivery problems: A classification scheme and
  survey.
\newblock {\em TOP}, 15(1):1--31, July 2007.

\bibitem{berbeglia_dynamic_2010}
Gerardo Berbeglia, Jean-Fran{\c c}ois Cordeau, and Gilbert Laporte.
\newblock Dynamic pickup and delivery problems.
\newblock {\em European Journal of Operational Research}, 202(1):8--15, April
  2010.

\bibitem{baldacci_exact_2011}
Roberto Baldacci, Enrico Bartolini, and Aristide Mingozzi.
\newblock An {{Exact Algorithm}} for the {{Pickup}} and {{Delivery Problem}}
  with {{Time Windows}}.
\newblock {\em Operations Research}, 59(2):414--426, April 2011.

\bibitem{grandinetti_multi-objective_2014}
L.~Grandinetti, F.~Guerriero, F.~Pezzella, and O.~Pisacane.
\newblock The {{Multi}}-objective {{Multi}}-vehicle {{Pickup}} and {{Delivery
  Problem}} with {{Time Windows}}.
\newblock {\em Procedia - Social and Behavioral Sciences}, 111:203--212,
  February 2014.

\bibitem{cordeau_dial--ride_2007}
Jean-Fran{\c c}ois Cordeau and Gilbert Laporte.
\newblock The dial-a-ride problem: Models and algorithms.
\newblock {\em Annals of Operations Research}, 153(1):29--46, September 2007.

\bibitem{parragh_hybrid_2013}
Sophie~N. Parragh and Verena Schmid.
\newblock Hybrid column generation and large neighborhood search for the
  dial-a-ride problem.
\newblock {\em Computers \& Operations Research}, 40(1):490--497, January 2013.

\bibitem{mahmoudi_finding_2016}
Monirehalsadat Mahmoudi and Xuesong Zhou.
\newblock Finding optimal solutions for vehicle routing problem with pickup and
  delivery services with time windows: {{A}} dynamic programming approach based
  on state-space-time network representations.
\newblock {\em Transportation Research Part B: Methodological}, 89(Supplement
  C):19--42, July 2016.

\bibitem{santi_quantifying_2014}
Paolo Santi, Giovanni Resta, Michael Szell, Stanislav Sobolevsky, Steven~H.
  Strogatz, and Carlo Ratti.
\newblock Quantifying the benefits of vehicle pooling with shareability
  networks.
\newblock {\em Proceedings of the National Academy of Sciences},
  111(37):13290--13294, September 2014.

\bibitem{tachet_scaling_2017}
R.~Tachet, O.~Sagarra, P.~Santi, G.~Resta, M.~Szell, S.~H. Strogatz, and
  C.~Ratti.
\newblock Scaling {{Law}} of {{Urban Ride Sharing}}.
\newblock {\em Scientific Reports}, 7:42868, March 2017.

\bibitem{alonso-mora_-demand_2017}
Javier {Alonso-Mora}, Samitha Samaranayake, Alex Wallar, Emilio Frazzoli, and
  Daniela Rus.
\newblock On-demand high-capacity ride-sharing via dynamic trip-vehicle
  assignment.
\newblock {\em Proceedings of the National Academy of Sciences},
  114(3):462--467, January 2017.

\bibitem{tanner_mathematical_1964}
J.~C. Tanner.
\newblock Mathematical {Theories} of {Traffic} {Flow}.
\newblock {\em Journal of the Operational Research Society}, 15(2):149--149,
  June 1964.

\bibitem{kerner_introduction_2009}
Boris~S. Kerner.
\newblock {\em Introduction to {Modern} {Traffic} {Flow} {Theory} and
  {Control}: {The} {Long} {Road} to {Three}-{Phase} {Traffic} {Theory}}.
\newblock Springer-Verlag, Berlin Heidelberg, 2009.

\bibitem{tadaki_critical_2015}
Shin-ichi Tadaki, Macoto Kikuchi, Minoru Fukui, Akihiro Nakayama, Katsuhiro
  Nishinari, Akihiro Shibata, Yuki Sugiyama, Taturu Yosida, and Satoshi Yukawa.
\newblock Critical {Density} of {Experimental} {Traffic} {Jam}.
\newblock In {\em Traffic and Granular Flow '13}, pages 505--511. Springer
  International Publishing, November 2015.

\bibitem{vcerticky2015fully}
Michal {\v{C}}ertick{\'y}, Jan Drchal, Marek Cuch{\'y}, and Michal Jakob.
\newblock Fully agent-based simulation model of multimodal mobility in european
  cities.
\newblock In {\em International Conference on Models and Technologies for
  Intelligent Transportation Systems (MT-ITS)}, pages 229--236. IEEE, 2015.

\bibitem{hensher2007handbook}
David~A Hensher and Kenneth~J Button.
\newblock {\em Handbook of transport modelling}.
\newblock Emerald Group Publishing Limited, 2007.

\bibitem{jass2016}
Jan Drchal, {\v{C}}ertick{\'y}, and Michal Jakob.
\newblock {VALFRAM: Validation Framework for Activity-Based Models}.
\newblock {\em Journal of Artificial Societies and Social Simulation},
  19(3):1--5, 2016.

\bibitem{drchal2015data}
Jan Drchal, Michal {\v{C}}ertick{\'y}, and Michal Jakob.
\newblock Data driven validation framework for multi-agent activity-based
  models.
\newblock In {\em International Workshop on Multi-Agent Systems and Agent-Based
  Simulation}, pages 55--67. Springer, 2015.

\bibitem{pavone_robotic_2012}
Marco Pavone, Stephen~L Smith, Emilio Frazzoli, and Daniela Rus.
\newblock Robotic load balancing for mobility-on-demand systems.
\newblock {\em The International Journal of Robotics Research}, 31(7):839--854,
  June 2012.

\bibitem{parragh_survey_2008}
Sophie~N. Parragh, Karl~F. Doerner, and Richard~F. Hartl.
\newblock A survey on pickup and delivery problems.
\newblock {\em Journal f{\"u}r Betriebswirtschaft}, 58(2):81--117, June 2008.

\bibitem{ropke_branch_2009}
Stefan Ropke and Jean-Fran{\c c}ois Cordeau.
\newblock Branch and {{Cut}} and {{Price}} for the {{Pickup}} and {{Delivery
  Problem}} with {{Time Windows}}.
\newblock {\em Transportation Science}, 43(3):267--286, June 2009.

\bibitem{cap_multi-objective_2018}
Michal {\v C}{\'a}p and Javier {Alonso-Mora}.
\newblock Multi-{{Objective Analysis}} of {{Ridesharing}} in {{Automated
  Mobility}}-on-{{Demand}}.
\newblock In {\em Proceedings of {{Robotics}}: {{Science}} and {{Systems}}},
  volume~14, 2018.

\bibitem{jung_dynamic_2015}
Jaeyoung Jung, R~Jayakrishnan, and Ji~Young~Park.
\newblock Dynamic {Shared}-{Taxi} {Dispatch} {Algorithm} with {Hybrid}
  {Simulated} {Annealing}.
\newblock {\em Computer-Aided Civil and Infrastructure Engineering}, 31, June
  2015.

\end{thebibliography}

\end{document}